\documentstyle[preprint,aps]{revtex}

\begin{document}
\tighten
\draft
\preprint{
 \parbox[t]{50mm}{hep-ph/9508323 
 \parbox[t]{50mm}{DPNU-95-20\\
}}}

\title{Confinement and complex singularities in QED3}
\author{P. Maris\cite{emmar}}
\address{Department of Physics, Nagoya University,
Nagoya 464-01, Japan}
\date{June 1995}
\maketitle
\begin{abstract}
The standard approximations of the Dyson--Schwinger equation lead to
complex singularities of the fermion propagator. In three-dimensional
QED one can show that this phenomenon might be related to confinement:
a confining potential leads to mass-like singularities at complex
momenta, and thus to the absence of a mass singularity on the real
timelike axis. The correct treatment of the vacuum polarization is
essential for the confining nature of QED3.
\end{abstract}

\section{Introduction}

Quarks are not observed as free particles, but only indirectly inside
hadrons; this confinement of the quarks is an essential and very
intriguing property of QCD, both from a theoretical and from an
experimental point of view. Despite a lot of effort, there are still a
lot of open questions about the confinement mechanism and confined
particles. One of such questions is what the behavior of the full
propagator of a confined particle is: e.g. does it have the same kind
of analyticity properties as a bare quark propagator? If the full
quark propagator has no mass singularity in the timelike region, it
can never be on mass-shell and thus never be observed as a free
particle \cite{Co80,GoMa89,Gr91,RoWiKr92}. So in this way the absence
of a mass singularity implies directly confinement, and thus the
analytic structure of the full quark propagator might be connected
with confinement.

Since confinement is a nonperturbative phenomenon, the analytic
properties of the full fermion propagator in a confining theory have
to be studied in a nonperturbative way. The Dyson--Schwinger equation
is a very powerful tool to study nonperturbative phenomena, and it is
commonly used for studying dynamical chiral symmetry breaking, but it
can also be useful in studies of confinement \cite{RoWi94}. The usual
truncation schemes of the Dyson--Schwinger equation show that the full
fermion propagator in QED and QCD has complex branchpoints, instead of
the expected mass singularity on the real timelike axis
\cite{AtBl79,StCa90,MaHo92,StCa92,Ma94}. Although this phenomenon
might be an artifact of the approximations, as believed about 15 years
ago when it was first discovered \cite{AtBl79}, it has been suggested
more recently that it might be a genuine property of the full theory,
connected with confinement, especially in QCD
\cite{StCa90,MaHo92,StCa92}. If the quark propagator has a mass-like
singularity at complex momenta, instead of a mass singularity in the
timelike region, it can never be on mass-shell and is thus confined.

Not only in QCD the fermions are confined, also in several other
theories there is confinement. Quantum electrodynamics in two space-
plus one time-dimension (QED3) is such a theory, with a confining
potential for the fermions, at least at the classical level; for the
full theory it depends on the behavior of the vacuum polarization
\cite{BuPrRo92}. It is also a very interesting model to study
dynamical mas generations, and for this purpose the Dyson--Schwinger
equation has been extensively studied on the Euclidean axis
\cite{Pi84,appetal,DaKoKo89,PeWe88,AtJoMa90,Penetal92}. The theory is
super-renormalizable, and does not suffer from the ultraviolet
divergences which are present in the corresponding four-dimensional
theories. That means that we do not need to introduce any artificial
cutoff, and the only mass scale in massless QED3 is the dimensionful
coupling. In this way we are provided with a very interesting model,
from which we can learn a lot about the analytic structure of the
propagator, and which is mathematically easier to analyze than
four-dimensional theories. The result can be very useful as guidance
for other, more complicated, theories like QCD. Apart from the the
interesting features connected with dynamical mass generation and
confinement in general, it might also have some direct physical
relevance, both in condensed matter physics (in connection with
phenomena occurring in planes) and as the high-temperature limit of
the corresponding four-dimensional theory.

In this paper we study the analytic structure of the fermion
propagator in QED3, using the Dyson--Schwinger equation and some
different approximations for the full photon propagator. We show that,
if there is a confining potential, the fermion propagator has complex
mass-like singularities, but if there is no confining potential, the
mass singularities are located almost on the real timelike axis, as we
would expect. The presence or absence of the confining potential
depends on the particular approximation for the photon propagator.

This paper is organized as follows: in the next section we review the
analytic structure of the fermion propagator in the context of
perturbation theory and what we would expect for the propagator of a
confined particle. In Sec.~\ref{secform} we introduce the model we are
considering and its confining properties. Next, we discuss the
Dyson--Schwinger equation, the truncation scheme we are using, and our
numerical procedures. In Sec.~\ref{secres} we present our results, and
finally we give some conclusions in Sec.~\ref{secconc}.

\section{Analytic structure of propagators}
\label{secans}

\subsection{Mass singularities}

The analytic structure of the bare fermion propagator is well-known:
in momentum space, it has a single pole at the bare mass of the
fermion. In Minkowski metric (which we will use in this section only,
out of convenience), we have for the bare propagator
\begin{eqnarray}
 S(p) &=&  \frac{1}{\not{\! p} - m_0 + i\varepsilon} \,,
\end{eqnarray}
with a mass-pole which is located at timelike momentum
$p^2_{\hbox{\scriptsize Mink}} = m^2_0$. The integration contour one
encounters in all kinds of calculations, goes around this singularity
due to the $i\varepsilon$-prescription, and this $i\varepsilon$ also
allows us to perform the usual Wick rotation from Minkowski space to
Euclidean space.

In perturbation theory, the full fermion propagator has a similar
structure, at least on the first Riemann sheet: a single pole at the
physical mass of the particle, and a more complicated structure for
momenta beyond some threshold energy for multi-particle production,
see Fig.~\ref{fig1}. If we are dealing with massless particles, as in
QED, where we have massless photons, this single pole becomes a
logarithmic branchpoint, see e.g. \cite{itzu}.

In general, we expect a similar structure for the full fermion
propagator in a nonperturbative calculation, at least if the fermion
corresponds to a stable physical particle. In a theory of interacting
particles {\em with asymptotic states} we have the K\"allen--Lehmann
representation
\begin{eqnarray}  \label{klrep}
 S_F(p) &=&  Z_2 \frac{\not{\! p} + m_{\hbox{\scriptsize phys}}}
         {p^2 - m_{\hbox{\scriptsize phys}}^2}
  +  \int_{\tilde m^2}^\infty {\rm d}\mu^2
     \frac{\not{\! p} \rho_1(\mu^2) + \rho_2(\mu^2)}
         {p^2 - \mu^2 + i\varepsilon} \,,
\end{eqnarray}
with the spectral weight functions $\rho_i(p^2)$ real and nonnegative,
and $\tilde m \ge m_{\hbox{\scriptsize phys}}$ the threshold for
multi-particle production.  We therefore expect a full electron
propagator with a mass singularity at the physical mass of the
electron, which is located on the real axis in the timelike region at
$p^2_{\hbox{\scriptsize Mink}} = m^2_{\hbox{\scriptsize phys}}$, and a
logarithmic branch-cut along the real axis, beyond this singularity.

However, the derivation of the K\"allen--Lehmann representation breaks
down in the absence of the asymptotic states; the above argument only
holds in cases were the fermion is indeed a stable, physically
observable particle. If we are considering a theory with confined
fermions, which means that there are no asymptotic states for these
fermions, we do not have a rigorous proof of the existence of a
K\"allen--Lehmann representation, so we do not know a priori the
analytic structure of the propagator of such a confined particle.

The mass singularities of the propagator at the physical mass of the
particle are crucial for the existence of observable asymptotic
states. Without such mass singularities, the particles can never be on
mass shell, and thus never be observed as real particles. In other
words, confinement might very well be related to the absence of such
mass singularities, and thus to the absence of a K\"allen--Lehmann
representation for the propagator of such a particle
\cite{Co80,GoMa89,Gr91,RoWiKr92}.

\subsection{Complex singularities}

If the propagator of a confined particle does not have a a
K\"allen--Lehmann representation, what (other) analytic structure can
we expect? Writing the full fermion propagator as
\begin{eqnarray}
     S(p) &=& Z(p^2)\frac{\not{\! p} + m(p^2)}{p^2 - m^2(p^2)} \,,
\end{eqnarray}
we can now ask the question: what analytic structure is possible?
In principle there are the following possibilities:
\begin{itemize}
 \item the propagator has complex singularities
 (at zeros of the denominator);
 \item the propagator is an entire function;
 \item the propagator has compensating zeros
 (both the denominator and the wavefunction renormalization $Z(p)$ are
 zero at the same point, which might be located in the timelike region).
\end{itemize}

During the last couple of years, analyses of the fermion propagator
using the Dyson--Schwinger equation in the complex momentum plane show
complex mass-like singularities in a variety of models and truncation
scheme. This phenomenon was first discovered by Atkinson and Blatt
\cite{AtBl79} in quenched ladder QED4, and it was generally believed
to be an artifact of the approximations. However, in a theory with
confined particles, it might very well be a genuine property of the
full theory: the absence of a mass singularity at timelike momenta
will effectively confining the particles, in the sense that they will
not be observable as physical stable states. Recently, it has been
suggested by several authors, that the complex singularities one finds
by solving the Dyson--Schwinger equation for complex momenta, are
indeed a signal for confinement, especially in a confining theory like
QCD \cite{StCa90,MaHo92,StCa92}. In this paper we show that there is
indeed a connection between a confining potential and complex
mass-like singularities in QED3.

Note however that also other analytic structures, like a fermion
propagator which is an entire function, will effectively confine the
fermions; and in principle there are also other confinement mechanisms
possible which do allow for a physical mass pole for the fermion
propagator.

\section{QED3}
\label{secform}

\subsection{Formalism}

In Minkowski space, we need the artificial $i\varepsilon$ description,
in order to define the path integrals, and to select the integration
path around the mass singularities in all kinds of calculations.
Alternatively, we could set up our field theory in Euclidean space, in
which case the integrals are well-defined from the beginning. In
principle the Wick rotation allows us to go from Euclidean to
Minkowski space and back, and both formulations seem to be equivalent,
but in the presence of complex singularities this easy connection
between Euclidean and Minkowski space is destroyed. Since the theory
is better defined in Euclidean metric, we will use that formalism.
Once we know the Euclidean Green's functions, we can obtain the
Wightman functions in coordinate space by an analytic continuation in
the time-coordinates, and from them the physically relevant Minkowski
Green's functions \cite{glijaf}. In this way we can (in principle)
extract all of the physically relevant information in Minkowski space,
even after setting up the formalism in Euclidean space.

The Lagrangian in Euclidean space is
\begin{eqnarray}
  {\cal L}(\psi, \bar\psi, A) &=&
 \bar\psi \left( \gamma^\mu(\partial^\mu + ieA^\mu) + m_0 \right) \psi
    + \textstyle{\frac{1}{4}} F^{\mu\nu}F^{\mu\nu}
     + \textstyle{\frac{1}{2a}} (\partial^\mu A^\mu)^2 \,.
\end{eqnarray}
In QED3, only three anti-commuting $\gamma$-matrices are needed, which
can be realized by taking a two-dimensional representation for these
matrices, e.g. the Pauli spin matrices. In that case we also have
two-dimensional spinors, instead of the four-dimensional spinors one
would use in four-dimensional theories. However, here we will use the
formulation with a four-dimensional spinor-space, and use the same
$\gamma$-matrices as in four space-time dimensions. In order to study
dynamical mass generations, and its influence on confinement, we take
the bare fermions to be massless: $m_0 = 0$. In QED3 with
four-dimensional spinors one can have two types of mass terms for the
fermions, namely a parity-breaking and a parity-conserving mass term.
We will only consider the dynamical generation of a parity-even mass
\cite{appetal}; it has been shown that there is no dynamical breakdown
of parity \cite{VaWi84}.

The full fermion propagator can be written as
\begin{eqnarray}
  S^{-1}(p) &=& Z(p) \left( i\not{\! p} + m(p) \right) \,,
\end{eqnarray}
and the full photon propagator is
\begin{eqnarray}   \label{photprop}
  D^{\mu\nu}(q) &=& \frac{1}{q^2(1 + \Pi(q))}
    \left( \delta^{\mu\nu} - \frac{q^\mu q^\nu}{q^2} \right)
        + a \frac{q^\mu q^\nu}{q^4} \,,
\end{eqnarray}
in a general covariant gauge. In this equation, $\Pi(q)$ is the vacuum
polarization, $a$ the gauge parameter, $m(p)$ the dynamical mass
function of the fermion, and $Z(p)$ the fermion wavefunction
renormalization. For sake of simplicity, we use the Landau gauge
($a=0$).

The exact Dyson--Schwinger equation for the fermion propagator is
\begin{eqnarray}     \label{dseqn}
  S^{-1}(p) &=& i\not{\! p} + e^2 \int\frac{{\rm d}^3k}{(2\pi)^3}
             \gamma^\mu S(p) \Gamma^\nu(p,k) D^{\mu\nu}(p-k) \,,
\end{eqnarray}
with the unknown full vertex $\Gamma^\nu(p,k)$, and the full photon
propagator $D_{\mu\nu}(p-k)$. In analyzing the fermion
Dyson--Schwinger equation, we have to truncate this equation. In this
paper, we discuss both the socalled quenched ladder approximation
(bare photon and bare vertex), and two approximations based on the
$1/N$ expansion.

\subsection{$1/N$ Expansion}

A very popular truncation scheme in QED3 is the $1/N$ expansion:
consider $N$ massless fermion flavors, and use the large $N$ limit in
the following way: let $N \rightarrow \infty$ and $e^2 \rightarrow 0$
in such a way that the product $e^2 N$ remains fixed. It has been
shown that massless QED3 is infrared finite order by order in such a
$1/N$ expansion \cite{expansion}. For convenience we choose the
coupling, which defines our mass scale, to be
\begin{eqnarray}
  e^2 &=& \frac{8 \alpha}{N} \,,
\end{eqnarray}
and keep $\alpha$ fixed.

Such a $1/N$ expansion means that we have to take into account the
one-loop vacuum polarization: the coupling is of order $1/N$, but
there are $N$ fermion loops contributing to the vacuum polarization
tensor
\begin{eqnarray}  \label{vacpol}
  \Pi^{\mu\nu}(q) &=& - e^2 \, N  \int\frac{{\rm d}^3k}{(2\pi)^3}
  {\rm Tr}\left[ \gamma^\mu S(k+q) \Gamma^\nu(k+q,k) S(k) \right] \,.
\end{eqnarray}
This vacuum polarization tensor has an ultraviolet divergence in its
part proportional to $\delta^{\mu\nu}$, which can be removed by
evaluating it using a gauge-invariant regularization scheme. Defining
the vacuum polarization $\Pi(q)$ by
\begin{eqnarray}
  \Pi^{\mu\nu}(q) &=&
  \left(q^2 \delta^{\mu\nu} - q^\mu q^\nu \right) \Pi(q) \,,
\end{eqnarray}
we can get the regularized vacuum polarization by contracting the
vacuum polarization tensor with \cite{RoWi94,BuPrRo92}
\begin{eqnarray}
  \left( q^2 \delta^{\mu\nu} - 3 q^\mu q^\nu \right) /q^4 \,,
\end{eqnarray}
which is orthogonal to $\delta^{\mu\nu}$ and thus projects out the
divergent part. Note that this regularization gives the same result as
dimensional regularization, but this projection is much easier to
perform if we take into account dynamical fermions, in
Sec.~\ref{subsecdyn}. Using bare, massless fermions and a bare vertex,
we have for this vacuum polarization
\begin{eqnarray}  \label{vacpolml}
  \Pi(q) &=& \frac{e^2 \, N}{8 \sqrt{q^2}} \;=\; \frac{\alpha}{q} \,,
\end{eqnarray}
whereas the one-loop vacuum polarization with massive fermions gives
\begin{eqnarray}  \label{vacpolmass}
  \Pi(q) &=& \frac{2\alpha}{\pi \, q^2}
      \left( 2m + \frac{q^2 - 4m^2}{q}
           \arcsin{\frac{q}{\sqrt{q^2 + 4m^2}}} \right) \,.
\end{eqnarray}
The crucial difference between the vacuum polarization with massless
and with massive fermions lies in the infrared region: with massless
fermions the vacuum polarization blows up at the origin, $\Pi(0)
\rightarrow \infty$. With massive fermions however, with a constant
mass $m$, the vacuum polarization is finite in the infrared
\begin{eqnarray}   \label{vacpolmassir}
  \Pi(0) &\rightarrow& \frac{4\alpha}{3\pi \, m}  \,.
\end{eqnarray}

\subsection{Confining potential}

One of the interesting properties of QED3 is that it exhibits
confinement \cite{GoMa82}. We can define a ``classical'' potential for
the fermions in coordinate space \cite{BuPrRo92}
\begin{eqnarray}
  V(\vec{x}) &=& -e^2 \int\frac{{\rm d}^2q}{(2\pi)^2}
   e^{i\vec{q}\cdot\vec{x}} \frac{1}{\vec{q}^2(1+\Pi(\vec{q}^2))} \,,
\end{eqnarray}
where $\Pi(q)$ is the vacuum polarization. In lowest order in
perturbation theory, we can neglect the effects of the vacuum
polarization and simply calculate the potential. This leads to a
logarithmically rising potential
\begin{eqnarray}
    V(\vec{x}) &=& \frac{e^2}{2\pi} \ln(e^2 |x|) \,.
\end{eqnarray}
Because this potential increases at large distances, it effectively
confines the fermions, and there are no free asymptotic one-fermion
states possible.

Of course, this potential will change under the influence of the
vacuum polarization. As our results show, the correct inclusion of the
vacuum polarization is indeed essential for confinement. The relevant
region for the question whether or not there is a confining potential
is the infrared region, corresponding to large spatial separations in
configuration space. As is shown by Burden {\em et al.}
\cite{BuPrRo92}, under quite general and natural conditions for the
vacuum polarization, the potential associated with the full photon
propagator behaves like
\begin{eqnarray}
 V(\vec{x}) &=& \frac{e^2\,\ln{\big(e^2 |x|\big)} }{(1 + \Pi(0))\,2\pi}
               + \hbox{constant} + {\cal O}\left(\frac{1}{|x|}\right) \,.
\end{eqnarray}
This behavior can be derived assuming that the vacuum polarization is
bounded and continuously differentiable for Euclidean momenta, and that
it falls of at least as $1/q$ as $q \rightarrow \infty$.

{}From this equation, we can see immediately that, depending on the
behavior of the vacuum polarization in the infrared region, there are
two possibilities
\begin{itemize}
 \item a confining potential if $\Pi(0)$ is finite;
 \item no confining potential if $\Pi(0) \rightarrow \infty$.
\end{itemize}
If we know look at the one-loop perturbative vacuum polarization, we
see that there is an essential difference between massless and massive
QED3. For massive fermions, the vacuum polarization at the origin,
$\Pi(0)$, is finite as can be seen from Eq.~(\ref{vacpolmassir}).
Therefore there is a logarithmically confining potential in leading
order in $1/N$ for massive fermions.

On the other hand, if the fermions are massless, there is no confining
potential to leading order to $1/N$; the one-loop vacuum polarization
blows up at $q^2 \downarrow 0$, see Eq.~(\ref{vacpolml}), and the
photon propagator is softened in the infrared region. Perturbatively,
to leading order in $1/N$, the fermion propagator is just the bare
propagator, with a single pole at the origin, corresponding to an
observable massless fermion. However, a dynamically generated fermion
mass might very well change this leading-order behavior.

\section{Dyson--Schwinger equation}
\label{secdse}

In order to determine the analytic structure of the fermion propagator
nonperturbatively, we use the Dyson--Schwinger equation. In general,
after reducing the $\gamma$-algebra, the Dyson--Schwinger equation,
Eq.~(\ref{dseqn}), becomes
\begin{eqnarray}   \label{dsmassgen}
  Z^{-1}(p)\,m(p) &=& e^2 \int\frac{{\rm d}^3k}{(2\pi)^3}
      \textstyle{\frac{1}{4}}
 {\rm Tr}[\gamma^\mu S(k) \Gamma^\nu(p,k) D^{\mu\nu}(q)] \,,
\\ %
\label{dswavegen}
  Z^{-1}(p) & = & 1 - \frac{e^2}{p^2} \int\frac{{\rm d}^3k}{(2\pi)^3}
      \textstyle{\frac{1}{4}}
 {\rm Tr}[\not\! p \gamma^\mu S(k) \Gamma^\nu(p,k) D^{\mu\nu}(q)] \,,
\end{eqnarray}
with the photon propagator $D^{\mu\nu}(q)$ as defined by
Eq.~(\ref{photprop}), with the unknown vacuum polarization, and the
unknown full vertex function $\Gamma^\nu(p,k)$. The general approach
to solve this equation is to choose a specific truncation scheme for
the vertex and the photon propagator, and then to solve the resulting
equations numerically.

\subsection{Truncation scheme}

Here, we will adopt the bare vertex approximation, replacing the full
vertex by the bare one, $\gamma^\mu$. We also neglect the effects of
the wavefunction renormalization, so we put $Z(p) = 1$. This
truncation is based on the leading-order behavior of the vertex and
the wavefunction renormalization in the $1/N$ expansion.  Such an
approximation scheme is also consistent with the requirement following
from the Ward--Takahashi identity that the wavefunction
renormalization and the vertex renormalization are equal. It is
usually referred to as the ladder or rainbow approximation, and it
leads to a finite critical number of fermion flavors below which the
chiral symmetry is broken dynamically \footnote{Using this
approximation, the equation for the wavefunction renormalization,
Eq.~(\ref{dswavegen}), is formally satisfied up to order $1/N$. It is
known that the effects of the wavefunction renormalization (together
with a more sophisticated Ansatz for the vertex) will change the
results found in this $1/N$ truncation scheme
\cite{PeWe88,AtJoMa90,Penetal92}, but we will not address that problem
here.}. Note that in the Landau gauge, with a bare photon propagator
and bare vertex (the quenched ladder approximation),
Eq.~(\ref{dswavegen}) gives $Z(p)=1$ exactly.

For the photon propagator we use some different approximations, to
determine the influence of the infrared behavior of the vacuum
polarization on the analytic structure of the fermion propagator and
on the (confining) potential. We compare in detail the results as
obtained in
\begin{enumerate}
 \item the quenched approximation: a bare photon propagator;
 \label{trunc1}
 \item the $1/N$ expansion using the analytical formula for
  the one-loop vacuum polarization with bare, massless fermions,
  Eq.~(\ref{vacpolml})
 \label{trunc2}
 \item the $1/N$ expansion using the one-loop vacuum polarization
  with full fermions, {\em with the dynamically generated fermion
  mass function}.
 \label{trunc3}
\end{enumerate}
In both \ref{trunc2} and \ref{trunc3}, we take a bare vertex in the
expression for the vacuum polarization.

This truncation scheme gives us the following expression for the mass
function
\begin{eqnarray}    \label{genmasseq}
  m(p) &=& \frac{e^2}{2\,\pi^2} \int_0^\infty {\rm d}k
    \frac{k^2 \, m(k)}{k^2 + m^2(k)} {\rm K}(p,k) \,,
\\ %
  {\rm K}(p,k) &=& \int_0^\pi \frac{\sin{\theta} \, {\rm d}\theta}
     {(p^2-2pk\cos{\theta}+k^2)(1+\Pi(p^2-2pk\cos{\theta}+k^2))} \,,
\end{eqnarray}
with the kernel ${\rm K}(p,k)$ depending on the particular
approximation we use for the photon propagator.

\subsection{Numerical calculations}

Once we have truncated the equations, we can solve the resulting
integral equation for the mass function numerically. We start by
solving the equation for Euclidean momenta $0 \leq p^2 < \infty$.
However, we are not really interested in the result on the Euclidean
axis (for a more detailed discussion about the existence of a critical
number of fermion flavors for dynamical chiral symmetry breaking we
refer to the literature
\cite{appetal,DaKoKo89,PeWe88,AtJoMa90,Penetal92}), but we want to
know the behavior of the propagator in the complex momentum plane. For
that purpose we have used two different approaches: one is a direct
analytic continuation of the integral equation, Eq.~(\ref{genmasseq}),
into the complex plane. This can be done by deforming the integration
contour and solving the integral equation along this new contour. Note
that it is not possible to keep the integration variable $k$ real, and
take only the external variable $p$ complex (after solving the
integral equation on the real axis), because of the analytic structure
of the kernel ${\rm K}(p,k)$. With massless photons, and thus a photon
propagator which has a singularity at the origin, there is a pinch
singularity at $p=k$, and we are forced to integrate through the point
$p=k$. So for complex momenta $p$ we have to solve the integral
equation along a deformed contour in the complex plane.

In practice, we change the integration contour by rotating it in the
complex plane, multiplying both the internal and the external variable
by a phase factor ${\rm e}^{i\phi}$, so we get the complex variables
$\tilde k = {\rm e}^{i\phi} k$ and $\tilde p = {\rm e}^{i\phi} p$, see
Fig.~\ref{fig2}. Since in QED3 the integral falls off rapidly enough
in the ultraviolet, there is no need to take into account the
contribution coming from the arc at infinity, in contrast to theories
with a finite cutoff like QED4. This procedure works quite well, until
one comes close to a singularity caused by a zero of the denominator
of the integration kernel
\begin{eqnarray}
    \frac{1}{k^2 + m^2(k)} \,,
\end{eqnarray}
where the numerical integration procedure becomes unstable. The
location in the complex plane of the actual singularity itself can be
obtained by extrapolating the numerical results to the ``physical''
mass $\mu$, defined by the zero of this denominator
\begin{eqnarray}
    - \mu^2 + m^2(\sqrt{-\mu^2}) &=& 0 \,.
\end{eqnarray}
For more details about our numerical procedure and the analytic
continuation, we refer to \cite{pmthesis}. In this way we can in
principle find the singularities in the complex plane, but it is a
very time-consuming numerical process, and does not always converge to
a stable solution.

Therefore we also used another method, based on the Euclidean-time
Schwinger function, to determine whether or not the propagator
corresponds to a physical observable state \cite{RoWi94,HoRoMc92}.
We define
\begin{eqnarray}
  \sigma(p^2) &=& \frac{m(p)}{p^2 + m^2(p)} \,,
\\
  \Delta(t) &=& \int{\rm d}^2 x \int \frac{{\rm d}^3p}{(2\pi)^3}
    {\rm e}^{i(p_3 t + \vec{p}\cdot\vec{x})} \sigma(p^2) \,.
\end{eqnarray}
Using this Schwinger function, one can show that if there is a stable
asymptotic state associated with this propagator, with a mass $m$,
then
\begin{eqnarray}
  \Delta(t) &\sim& {\rm e}^{- m t}
\end{eqnarray}
for large (Euclidean) $t$, so for the logarithmic derivative we get
\begin{eqnarray}
  \lim_{t \rightarrow\infty} \frac{\rm d}{{\rm d}t}
   \ln\left(\Delta(t)\right)  &=& - m \,,
\end{eqnarray}
whereas two complex conjugate mass-like singularities, with complex
masses $\mu = a \pm i\,b$, lead to an oscillating behavior like
\begin{eqnarray}
  \Delta(t) &\sim& {\rm e}^{- a t} \cos{(bt + \delta)}
\end{eqnarray}
for large $t$. This method is much less time-consuming to see whether
or not the propagator has a real mass-singularity or not, but it is
less accurate than solving the Dyson--Schwinger equation for complex
momenta in determining the (complex) mass-singularities.

\section{Results}
\label{secres}

\subsection{Quenched QED3}
\label{subsecquen}

In massless quenched QED3, there is no free parameter: the coupling in
QED3 is dimensionful and thus defines the energy scale, and there are
no other parameters. By choosing the Landau gauge, we satisfy the
requirement that the wavefunction renormalization and the vertex
renormalization are exactly equal: from Eq.~(\ref{dswavegen}) it
follows directly that in quenched QED using the Landau gauge,
$Z(p)=1$, and the equation for the mass function reduces to
\begin{eqnarray}
  m(p) &=& \frac{e^2}{2\pi^2} \int_0^\infty {\rm d}k
    \frac{m(k)}{k^2 + m^2(k)} \frac{k}{2p} \ln{\frac{(p+k)^2}{(p-k)^2}} \,.
\end{eqnarray}
Solving this equation on the Euclidean axis shows that there is
dynamical mass generation in this case, and the infrared mass $m(0)$
is proportional to the dimensionful coupling, as expected.

Next, we have calculated the Schwinger function, using the mass
function on the Euclidean axis, see Fig.~\ref{fig3}. This figure
clearly shows that there is no stable asymptotic one-fermion state
associated with this propagator, in other words, the fermions cannot
be observed as free particles and are thus confined. The oscillations
in this figure strongly suggest that the fermion propagator has
complex mass-like singularities, corresponding to two
complex-conjugate masses. Using
\begin{eqnarray}  \label{schwcmplxmass}
  \Delta(t) &\sim& {\rm e}^{- a t} \cos{(bt + \delta)} \,,
\end{eqnarray}
to extract such a complex mass, we estimate this to be
\begin{eqnarray}
  \mu &=& (0.80 \pm 0.71 \, i)\,m(0)
\end{eqnarray}
However, this method might be not very accurate in determining the
actual value of the (complex) masses, since Eq.~(\ref{schwcmplxmass})
only holds for large values of $t$, whereas for large values of $t$
the numerical noise in calculating the Euclidean-time Schwinger
function destroys the signal. So we have used values of $t$ up to $t
\sim 10/m(0)$ (we rescale all dimensionful quantities by $m(0)$), and
use only the first oscillations to determine the imaginary part of the
complex mass. Furthermore, Eq.~(\ref{schwcmplxmass}) is based on the
assumption that only these (complex) singularities contribute to the
Schwinger function (at least at large $t$), but if there are complex
singularities, there might be more than just two complex-conjugate
mass-like singularities.

Therefore we also used the other method to determine the analytic
structure of the propagator, and have solved the integral equation in
the complex plane. This leads to two complex-conjugate singularities,
located at
\begin{eqnarray}
  |\mu| &=& 0.104 \,e^2 \;=\; 1.01\,m(0) \,,
\\
  \theta_\mu &=& {\textstyle{\frac{\pi}{2}}} - \phi \; = \; 0.819 \,.
\end{eqnarray}
This result confirms the estimate based on the Schwinger function,
given the inaccuracy of the estimate of the complex mass.

So both the Schwinger function and a direct search for mass-like
singularities show that there is a complex mass singularity, which
makes it impossible for the fermion propagator to become on
mass-shell, end thus effectively confines the fermion. This is in
agreement with the fact that in quenched massless QED3 there is a
confining potential.

\subsection{One-loop vacuum polarization}
\label{subsecbare}

Next, we include the one-loop vacuum polarization, using bare massless
fermions, Eq.~(\ref{vacpolml}). As already mentioned before,
perturbatively the $1/N$ expansion gives to leading order no confining
potential, and a full fermion propagator which is the same as the bare
one, and thus corresponding to a massless stable asymptotic state.  In
the case of dynamical mass generation, which we consider here, the
situation is more complicated. Since the number of fermion flavors is
the only free parameter in this case, we present our results as a
function of $N$.

For simplicity we use Landau gauge, as in the quenched approximation,
and we can perform the angular integration in the Dyson--Schwinger
equation analytically to arrive at the equation for the mass function
\begin{eqnarray} \label{masseqnml}
  m(p) &=& \frac{4\,\alpha}{N\,\pi^2}  \int_0^\infty {\rm d}k
            \frac{m(k)}{k^2 + m^2(k)} \frac{k}{p}
             \ln{\frac{|p+k|+\alpha}{|p-k|+\alpha}} \,,
\end{eqnarray}
This can be solved numerically as an integral equation, or after
expanding the logarithm and some further approximations reduced to a
second-order nonlinear differential equation \cite{appetal}. Both the
integral and the differential equation show that there is dynamical
mass generation if $N < N_c = 3.24$, see Fig.~\ref{fig4mass}; this
critical number is in agreement with analytical calculations using
bifurcation theory, leading to $N_c = 32/\pi^2$.

We have calculated the Schwinger function, using the mass function on
the Euclidean axis, see Fig.~\ref{fig3}. This figure strongly suggest that
there is a stable asymptotic one-fermion state, which means that the
fermions are not confined and can be observed.  Up to the largest
values of $t$ at which the Schwinger function gives a numerically
stable result, we find an almost constant logarithmic derivative.
{}From this Schwinger function we have derived a value for the
asymptotic mass for some different number of fermion flavors, and the
results are listed in Table~\ref{table}. We do not find any evidence
for oscillations which would signal a complex mass, as there were in
the quenched approximation.

We have also solved the integral equation in the complex plane. This
reveals that the mass singularities are not exactly on the real
timelike axis, but that they do have small imaginary parts, see
Table~\ref{table}. In Fig.~\ref{fig5phase} we have plotted the phase
of these singularities as a function of $N$, by numerically solving
both the integral equation and the differential equation which can be
derived from Eq.~(\ref{masseqnml}), and those results almost coincide.
Given the fact that we have only solved the truncated Dyson--Schwinger
equation, it is not unreasonable to expect a small deviation of the
physical mass from the real axis, and given the relative smallness of
the imaginary part this could very well be an artifact of the
approximations, especially since the singularity tends to move toward
the real timelike axis if the number of flavors goes to the critical
number.

The reason for not finding this imaginary part of the mass
singularities using the Schwinger function lies in its smallness:
since the imaginary part is of the order of $10\%$ of the real part
(or even less), we will not find a clear signal for it at $t \leq
10/m$, where $m$ is the typical infrared mass-scale; however, the
numerical noise destroys the signal completely at these (or larger)
values of $t$.

So our conclusion is that this approximation, using the one-loop
vacuum polarization of bare massless fermions, leads to (almost)
stable observable asymptotic states, with an (almost) real physical
mass. This agrees well with the fact that in this case we do not have
a confining potential.

\subsection{Full vacuum polarization}
\label{subsecdyn}

Finally, we include the one-loop vacuum polarization,
Eq.~(\ref{vacpol}), with dynamical fermions and a bare vertex. In other
words, we consider the coupled Dyson--Schwinger equations for the
photon and propagator, in the bare vertex approximation.  This leads
to two coupled integral equations to solve
\begin{eqnarray}
  m(p) &=& \frac{4\,\alpha}{N\,\pi^2}  \int_0^\infty {\rm d}k
            \frac{k^2 \, m(k)}{k^2 + m^2(k)}
  \int_{-1}^{1}\frac{{\rm d}z}{(p^2-2pkz+k^2)(1+\Pi(p^2-2pkz+k^2))} \,,
\\
  \Pi(q^2) &=&  \frac{8\alpha}{q^2} \int\frac{{\rm d}^3k}{(2\pi)^3}
  \frac{2 k^2 - 4 k\cdot q - 6 (k\cdot q)/q^2}
       {(k^2 + m^2(k))((k+q)^2 + m^2(k+q))}  \,.
\end{eqnarray}
Again, this can be solved numerically: we start by solving the mass
equation for a given vacuum polarization, and use that resulting mass
function to calculate the vacuum polarization numerically and iterate
this procedure. Just as in the previous case, it leads to dynamical
mass generation if the number of fermion flavors is below a critical
number, see Fig.~\ref{fig4mass}. The behavior of the infrared mass is
quite similar, and also the critical number is the same as in the
previous approximation, $N_c = 3.24$, as could be expected on grounds
of bifurcation theory.

Also in this case we have calculated the Schwinger function, see
Fig.~\ref{fig3extra}. This shows that there are no stable asymptotic
one-fermion states associated with this propagator, just as in
quenched QED3, but in sharp contrast to the previous case.  We have
also shown the result with a fixed mass ($m=m(0)$) in the analytical
formula for the vacuum polarization, Eq.~(\ref{vacpolmass}). This
gives qualitatively the same result as when using the dynamical mass
function. In both cases the oscillations indicate complex mass-like
singularities, and we have given estimates for these complex masses in
Table~\ref{table}.

We have also solved the integral equation in the complex plane, which
confirms the observation based on the Schwinger function that there
are complex mass-like singularities. The phase of these singularities
is plotted in Fig.~\ref{fig5phase}, and for some different values of
$N$ we have given our result in Table~\ref{table}. Given the
inaccuracy in the estimates based on the Schwinger function, there is
a good agreement between both methods.

It is clear that the the effects of the fermion mass in the loop for
the vacuum polarization confines the fermions: the potential becomes
confining, the mass singularities move into the complex plane, and
there are no stable observable asymptotic states.

\subsection{Discussion of the results}
\label{subsecdisc}

Our results show that the correct treatment of the vacuum polarization
is essential in a nonperturbative calculation of the fermion
propagator. Based on bifurcation theory, one can argue that the
influence of the dynamically generated mass function can be neglected
in studying the chiral phase transition. Although this is indeed true
for the value of the critical coupling, and maybe also for the
behavior of the infrared mass $m(0)$ close to the critical coupling,
it is certainly not true for the behavior of the {\em physical} mass,
defined at the zero of $p^2 + m^2(p)$.

On the real axis, the dynamical mass function in quenched QED3 is
qualitatively quite similar to the mass function in the $1/N$
expansion, both with massless fermions and with massive fermions in
the vacuum polarization, see Fig.~\ref{fig6}(a). There is a scale
difference between the different approximations, but all mass
functions are almost constant in the (far) infrared region, and fall
of to zero as $1/p^2$ in the (far) ultraviolet. Only in the
intermediate-energy region there are some differences due to the
inclusion of the vacuum polarization.

In contrast, in the complex plane the behavior is not similar at all,
leading to a drastic different analytic structure. This difference can
be traced back to the difference in the infrared behavior of the
photon propagator: with a confining photon propagator there are
complex mass-like singularities, whereas with a deconfining
photon propagator these singularities are located almost on the real
timelike axis. Surprisingly, this difference can be seen very clearly
by using the Euclidean-time Schwinger function, which can be
calculated using the mass function on the real Euclidean axis only.
So although the behavior of the mass function in the Euclidean region
looks quite similar, there are essential differences which can be
shown explicitly by calculating this Schwinger function. This means
that this method is indeed a useful way to determine whether the
propagator corresponds to a confined particle or to a physical
observable particle.

The difference in analytic structure is due to the difference in the
infrared behavior of the photon, or more precisely, due to the
different behavior of $\Pi(0)$ in the different approximations. In
Fig.~\ref{fig6}(b) we have plotted both the one-loop vacuum
polarization, for massless and massive fermions, and the full vacuum
polarization calculated numerically with dynamical massive fermions.
Only in the infrared region there is a difference, and it is exactly
this difference that causes the different behavior of the fermion
propagator in the complex plane; it is also this infrared behavior
which makes the potential confining or not. Therefore our conclusion
is that (at least in this model) confinement is caused by the infrared
behavior of the photon propagator, and is connected with complex
mass-like singularities of the fermion propagator, thus preventing the
fermions from being on mass-shell.

Finally, we should remark that these calculations are all done in the
bare vertex approximation in the Landau gauge. It is known that the
effects of vertex corrections, together with the wavefunction
renormalization $Z(p)$ which we have set equal to $1$, can change the
results quite drastically \cite{PeWe88,AtJoMa90,Penetal92}. Another
question is what happens in other gauges, whether or not our results
are gauge independent.  As a qualitative indication whether or not our
conclusions about confinement in QED3 also hold beyond the bare vertex
approximation and in other gauges, we could compare our numerical
vacuum polarization with the vacuum polarization as obtained by Burden
{\em et al}. They solved the Dyson--Schwinger equation for the fermion
propagator in quenched QED3 with the Ball--Chiu Ansatz for the vertex,
and used this propagator to calculate the vacuum polarization, again
with the Ball-Chiu vertex. Qualitatively our result for the vacuum
polarization agrees with theirs, both in the infrared region, where we
find a finite value of $\Pi(0)$ if we take into account the fermion
mass, and in the ultraviolet region. In the infrared region there is a
quantitative difference, but this can be explained by the fact that
the value of $\Pi(0)$ strongly depends on the infrared value of the
mass function $m(0)$, which is quite different in different
approximations. Of course, we should keep in mind that if the behavior
looks similar on the real axis, it does not necessarily mean that they
are indeed similar in the entire complex plane. However, the fact that
they also found a finite value of $\Pi(0)$ indicates that also beyond
the bare vertex approximation there is a confining potential, and we
would expect complex singularities as well. Whether or not these
singularities are gauge independent (with a suitable vertex Ansatz),
will be addressed in the future. Note that also the vacuum
polarization itself should be explicitly gauge-independent.

\section{Conclusions}
\label{secconc}

Our results show very clearly that there is a relation between a
confining potential, the absence of stable asymptotic states, and
complex mass-like singularities. Both in quenched QED3, and in
massive QED3 using the $1/N$ expansion (with a dynamically generated
fermion mass), there is a logarithmically confining potential, and we
show that there are no stable asymptotic states. The Euclidean-time
Schwinger function has an oscillatory behavior in these cases,
indicating complex mass-like singularities. By solving the
Dyson--Schwinger equation directly in the complex momentum plane, we
show that there are indeed such complex singularities.

On the other hand, using massless bare fermions in the one-loop vacuum
polarization, there is no confining potential. In this approximation,
the Schwinger function indicates a stable asymptotic state. A direct
analysis of the Dyson--Schwinger equation in the complex plane reveals
that there are complex singularities even in this case, but that they
are located very close to the real timelike axis. Given the
approximations made, it is not unreasonable to assume that this small
(less than $10\%$) deviation from the real timelike axis is caused by
the truncation of the Dyson--Schwinger equation.

These results are obtained with a dynamically generated fermion mass,
starting with massless bare fermions. For $N > N_c$, in the chirally
symmetric phase, there is no dynamical fermion mass, and the fermion
propagator has a singularity at the origin, just as the bare one. This
also agrees with the fact that in there is no confining potential in
this massless phase (in the $1/N$ expansion), due to the infrared
softening of the photon propagator in the presence of massless
fermions. Thus the chiral phase transition is a confining phase
transition as well, at least in this model.

Interpreting the absence of a mass singularity on the real axis in the
timelike region as confinement does not completely explain the
phenomenon of complex mass-like singularities. If they are indeed a
genuine property of the full theory, it leads automatically to
confinement, but it has more consequences. One of the consequences is
that the naive Wick rotation is not allowed, and one should take into
account the contributions coming from the complex singularities in
going from Euclidean to Minkowski metric (and back). Another problem
is connected with questions of unitarity an causality; however, one
should keep in mind that these are requirements for the S-matrix of
physical processes, and not necessarily for the propagator of an
unphysical (confined) particle.

Another question is whether or not these complex singularities have a
physical interpretation. Naively, the real part (or the absolute
value) could be interpreted as the ``constituent'' mass, and the
imaginary part as some ``hadronization length'', in terms of QCD and
quark confinement. Such an interpretation is analogous to the
interpretation of the poles of instable particles in terms of mass and
decay width. A crucial requirement for such an interpretation is that
the singularities are gauge independent, which has to be studied in
detail.

\section*{Acknowledgments}

I would like to thank Qing Wang, Craig Roberts, and David Atkinson
for useful comments.  This work has been financially supported by
the Japanese Society for the Promotion of Science.


\begin{figure}

\caption{The analytic structure of a full, stable,
physical observable particle, in (a) the $p_0$-plane and
(b) the $p^2$-plane (in Minkowski metric).}
\label{fig1}
\vspace{7mm}

\caption{The analytic continuation into the complex momentum
plane of the integral equation (in Euclidean momenta).}
\label{fig2}
\vspace{7mm}

\caption{The Schwinger function for quenched QED3 ($N=0$, diamonds)
and with the one-loop vacuum polarization ($N = 2$, plusses),
(a) the logarithmic derivative, (b) the logarithm of its absolute
value; the energy scale is defined by requiring $m(0) = 1$.}
\label{fig3}
\vspace{7mm}

\caption{The logarithm of the absolute value of the Schwinger function
with $N = 2$ and some different approximations for the photon propagator:
with the one-loop vacuum polarization of massless fermions
(Eq.~(\protect{\ref{vacpolml}}), plusses), with that of fermions
with a fixed mass $m(0)$ (Eq.~(\protect{\ref{vacpolmass}}), crosses),
and with the full vacuum polarization with dynamical fermions (squares);
for comparison, we also included the quenched results (diamonds).}
\label{fig3extra}
\vspace{7mm}

\caption{The infrared mass $m(0)$ with the one-loop vacuum polarization
of massless fermions (solid line) and with the full vacuum
polarization (diamonds), as a function of $N$}
\label{fig4mass}
\vspace{7mm}

\caption{The phase $\phi$ of the mass-singularity obtained
with the one-loop vacuum polarization of massless fermions
(both the results of using the integral equation
Eq.~(\protect{\ref{masseqnml}}) directly, and that of using a
differential equation which can be derived from it, after some
further approximations, see \protect{\cite{pmthesis}}),
and with the full vacuum polarization (diamonds), as a function of
$N$; for completeness we have included the value for quenched QED3
at $N=0$. The dotted line corresponds to $\phi = \pi/2$,
which means a mass singularity on the real timelike axis.}
\label{fig5phase}
\vspace{7mm}

\caption{The mass function (a) and vacuum polarization (b) as
obtained by taking into account the one-loop vacuum polarization of
massless fermions (dotted line), that of fermions with a fixed mass
$m(0)$ (dashed line), and that of dynamical fermions (solid line),
all for $N = 2$ and in units of $\alpha$; for completeness we included
the mass function for quenched QED3 (dashed-dotted line) in (a).}
\label{fig6}

\end{figure}

\begin{table}
\begin{tabular}{lllll}
 &\multicolumn{2}{l}{Schwinger function} &\multicolumn{2}{l}{Direct search}\\
 $N$ & $\hbox{Re}(\mu)$ & $\hbox{Im}(\mu)$ &
                                         $\hbox{Re}(\mu)$ & $\hbox{Im}(\mu)$\\
\hline
 \multicolumn{5}{l}{quenched QED3 ($\mu$ in units of $e^2$)}   \\
\hline
  0   & 0.082  & 0.073 & 0.0715 & 0.0760 \\
\hline
 \multicolumn{5}{l}{massless fermions in the vacuum polarization
($\mu$ in units of $\alpha$)}\\
\hline
 1   & 0.13    & 0   & 0.128   & 2.88e-2 \\
 2   & 2.6e-3  & 0   & 2.67e-3 & 1.4e-4 \\
 3   & 1.8e-9  & 0   & 1.9e-9  & 7e-11  \\
\hline
 \multicolumn{5}{l}{dynamical fermions in the vacuum polarization
($\mu$ in units of $\alpha$)}\\
\hline
 1  & 0.25    & 0.2    & 0.23   &  0.21  \\
 2  & 0.008   & 0.007  & 0.008  &  0.006 \\
\end{tabular}
\bigskip

\caption{
Estimates for the mass-singularities in the quenched approximation,
with massless fermions in the vacuum polarization and with the full
vacuum polarization, using both the Euclidean-time Schwinger function
and a direct search in the complex momentum plane.}
\label{table}

\end{table}

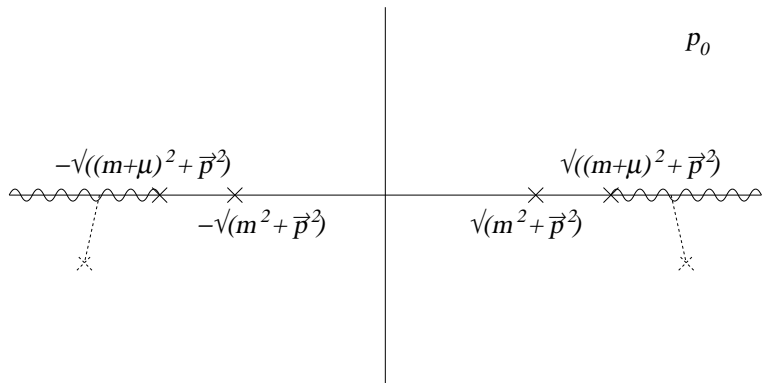
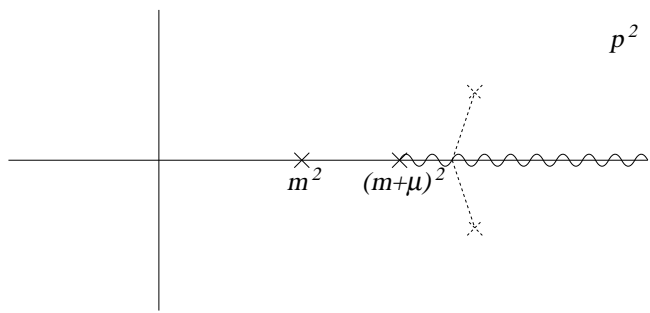
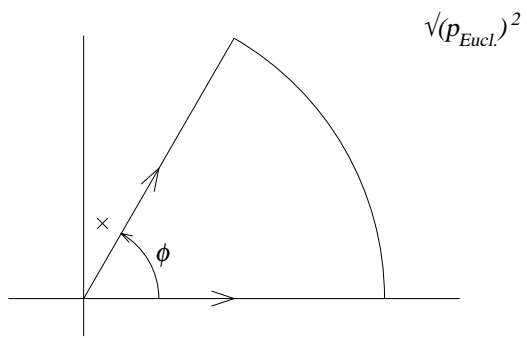
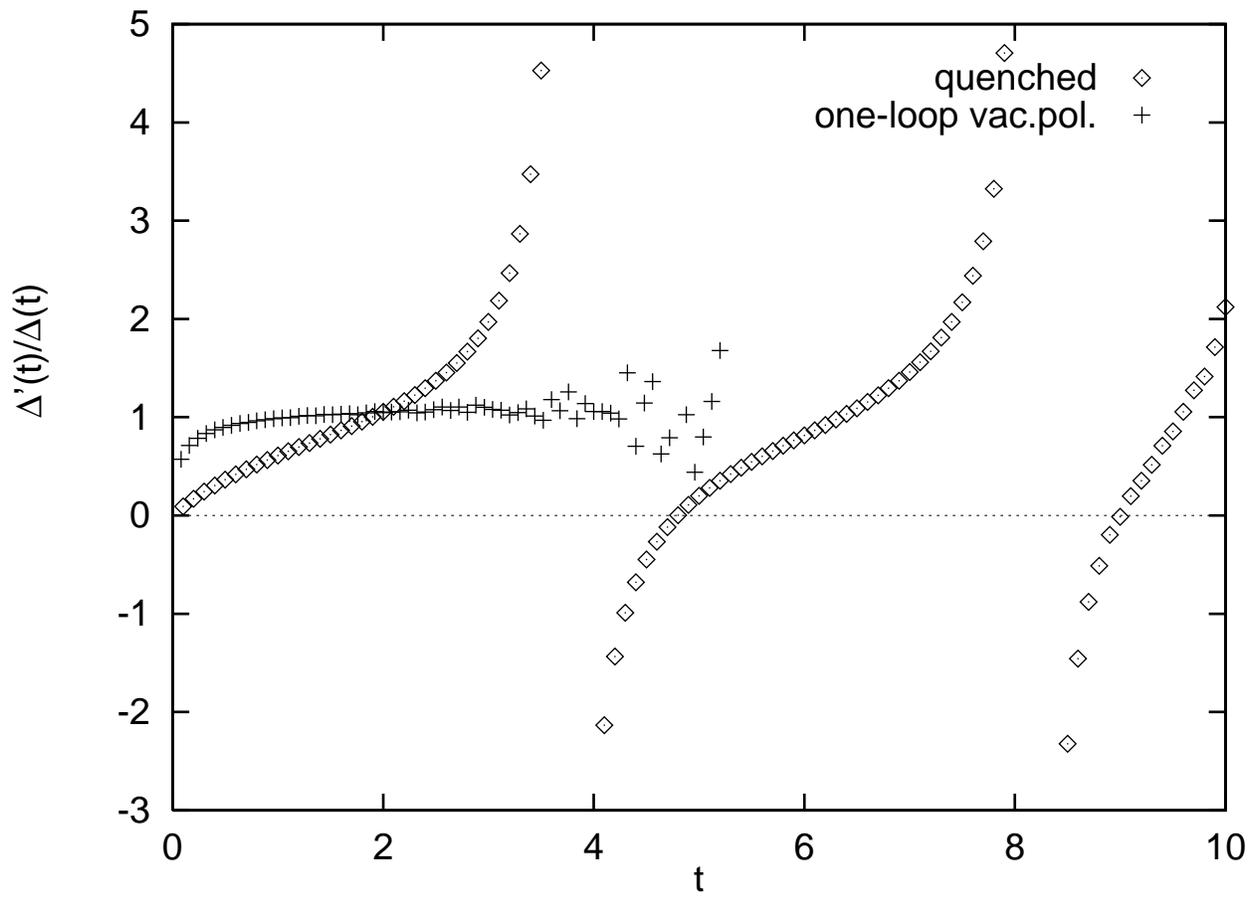
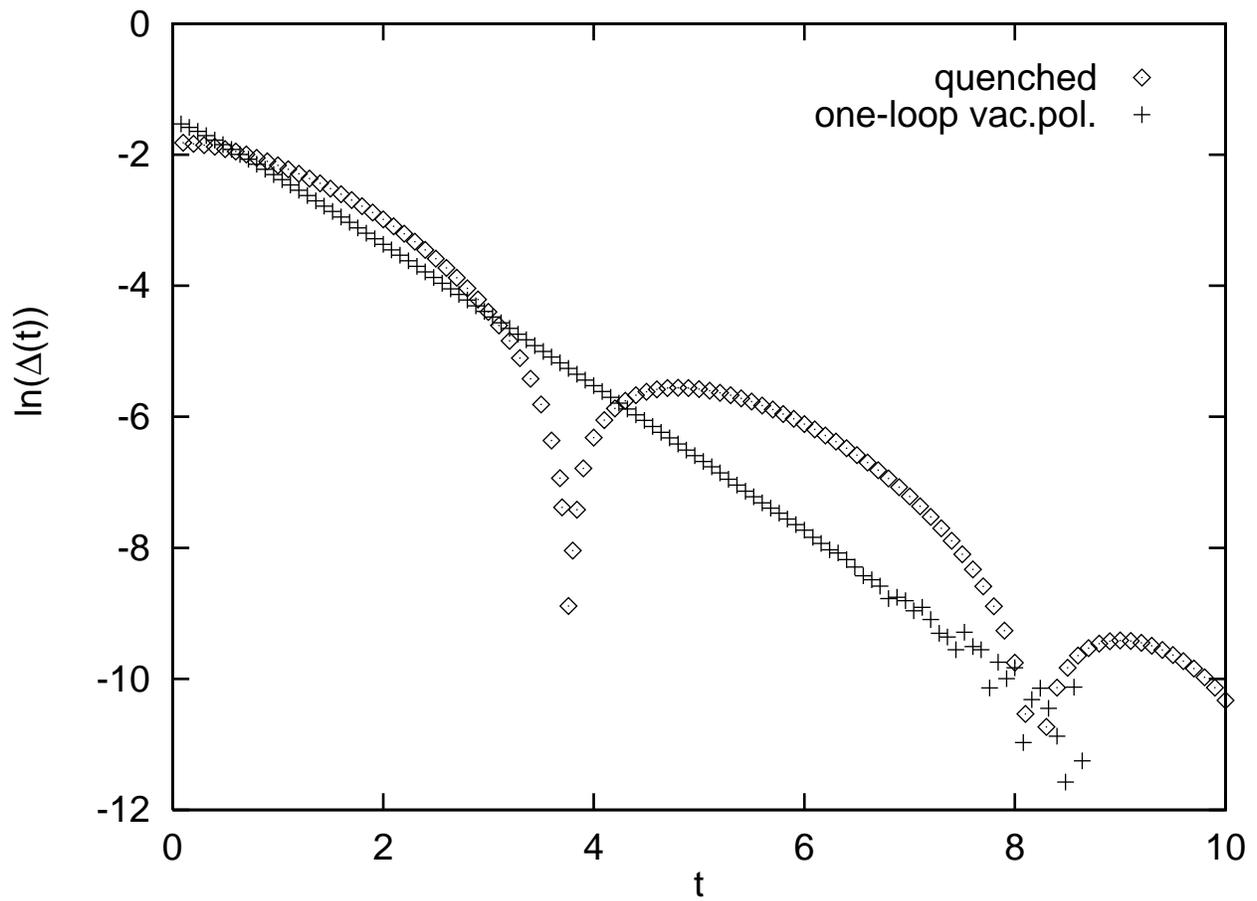
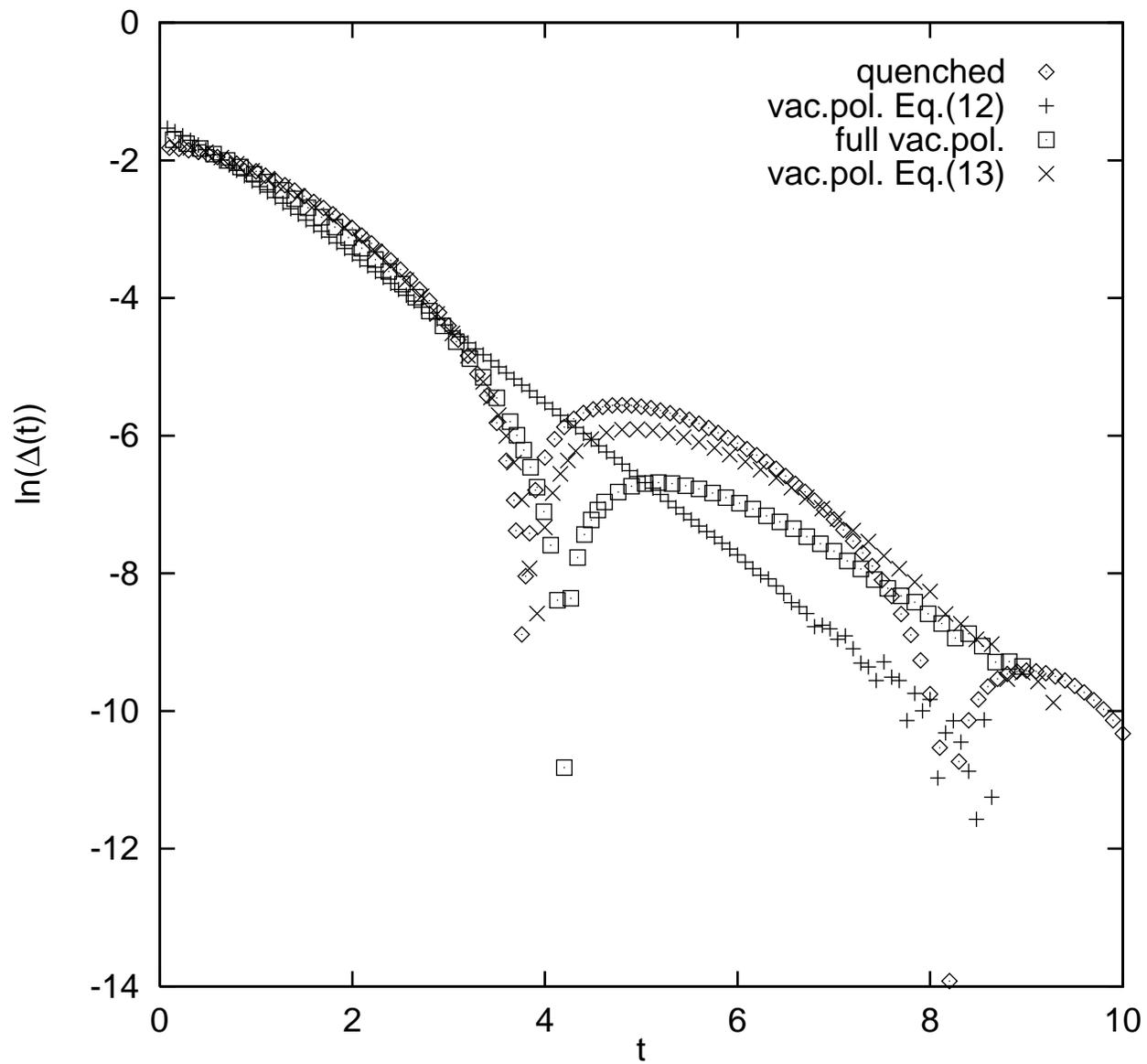
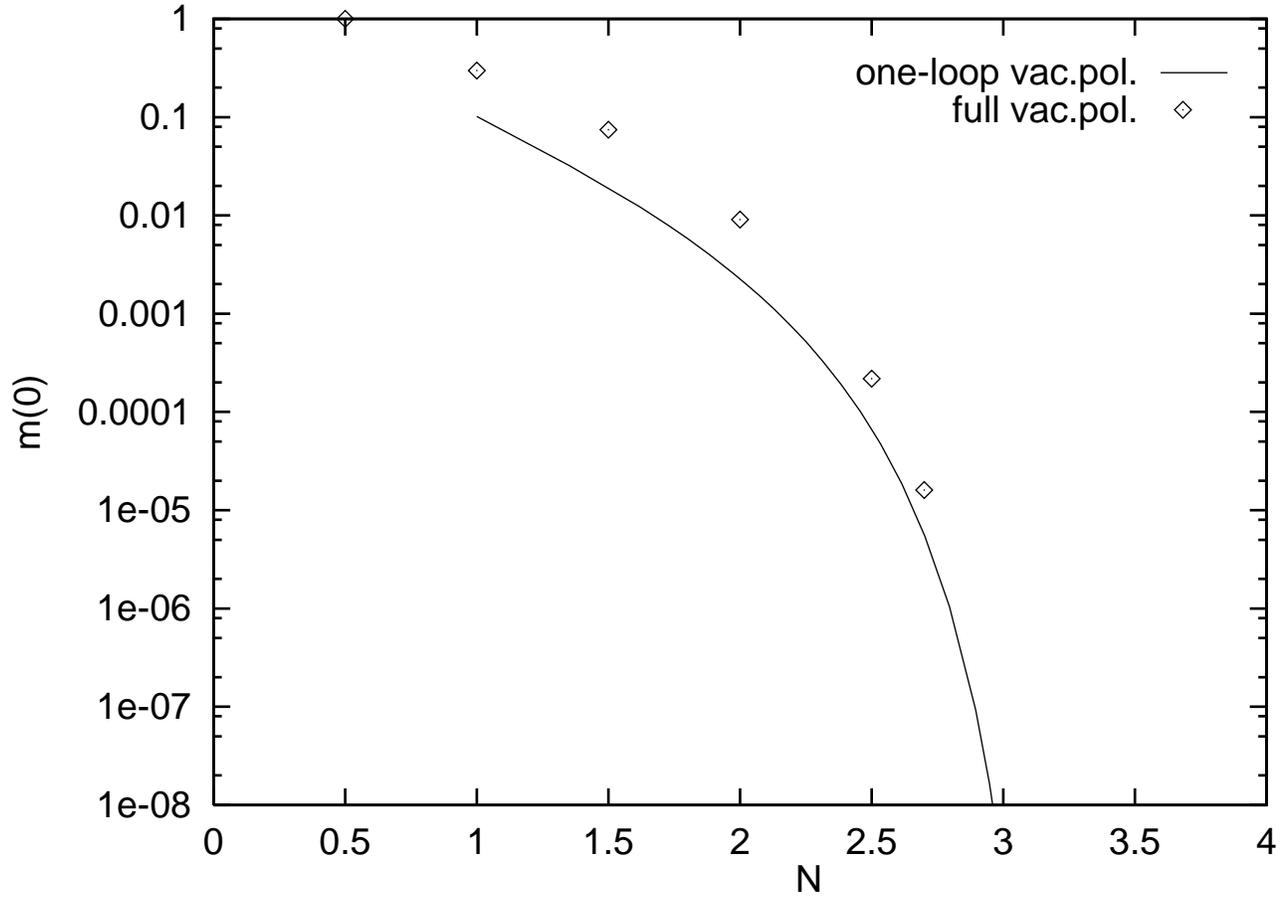
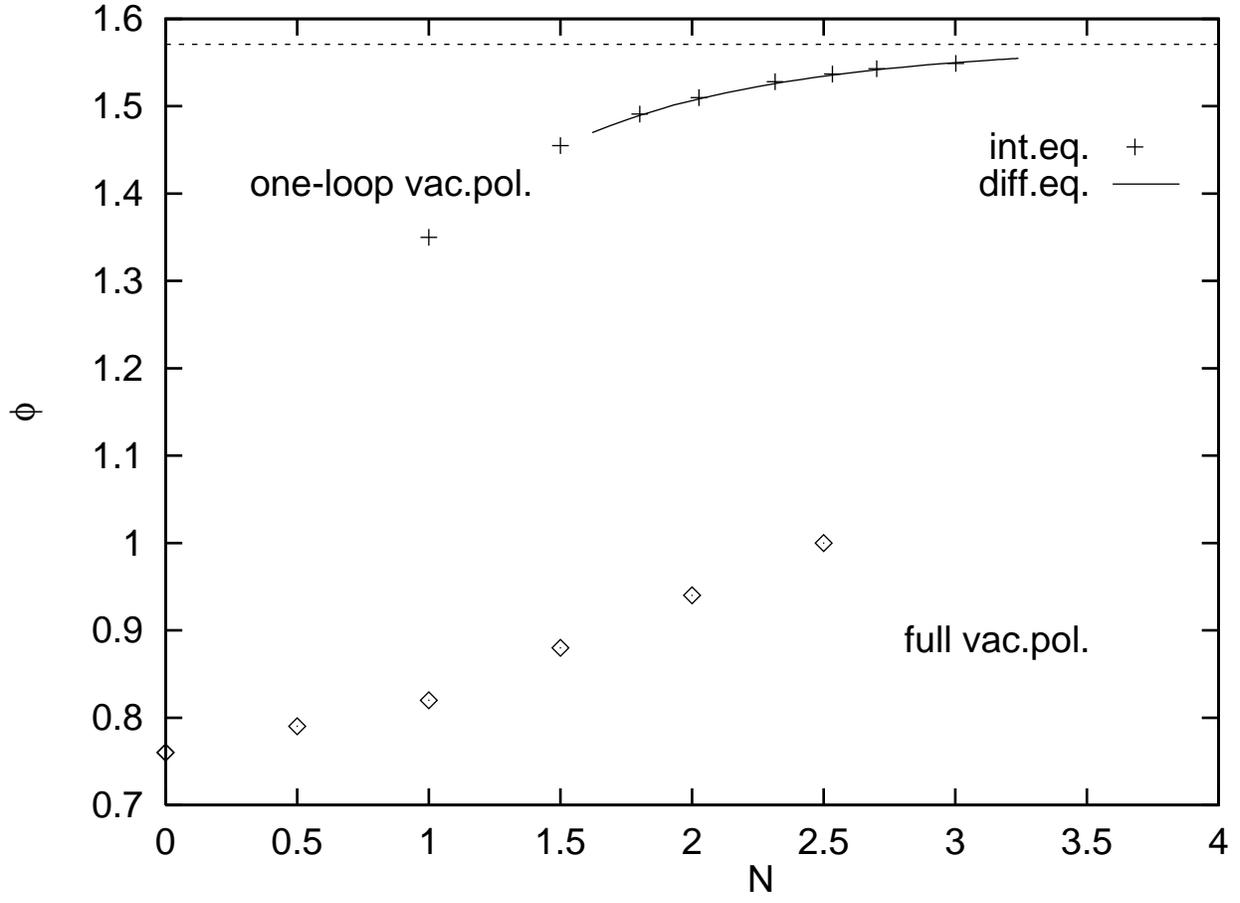
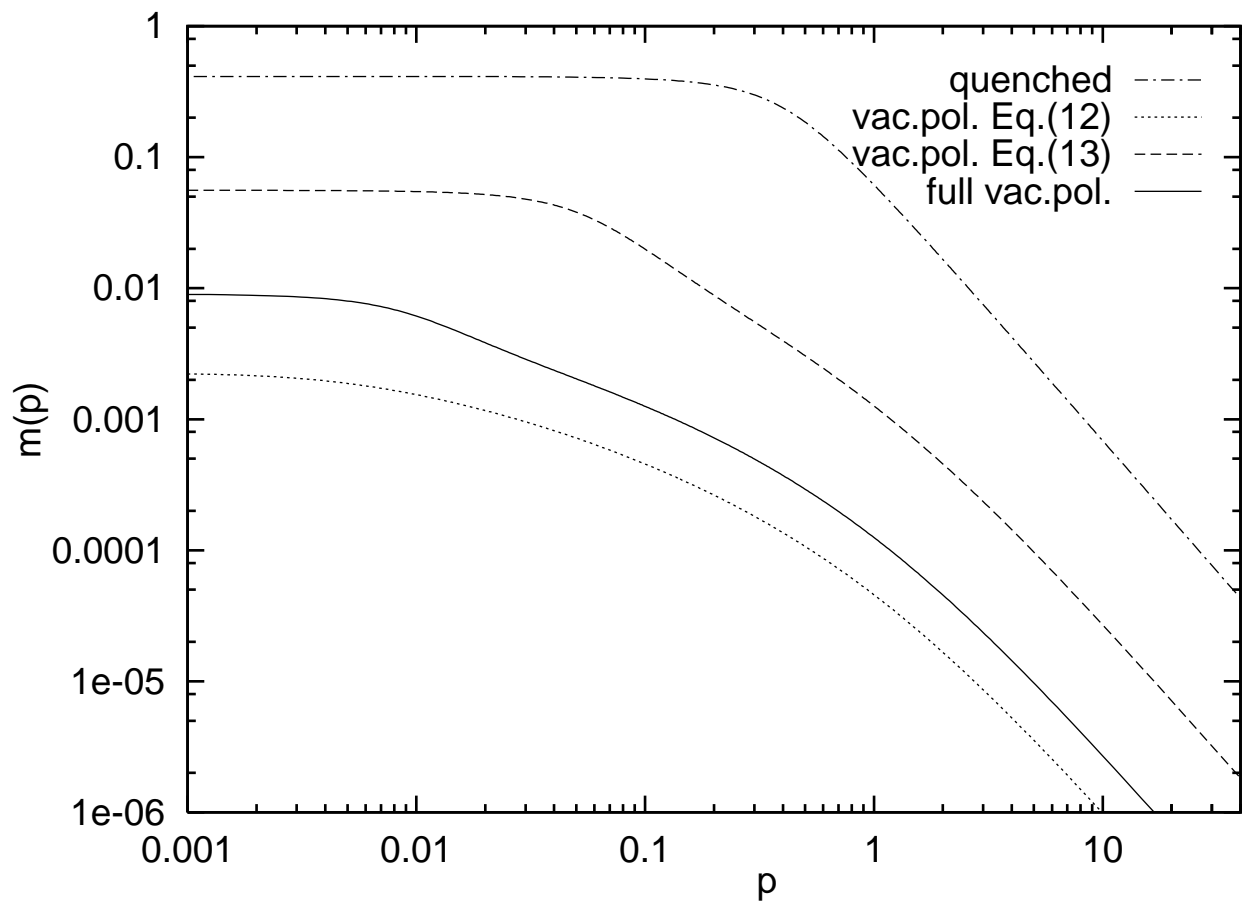
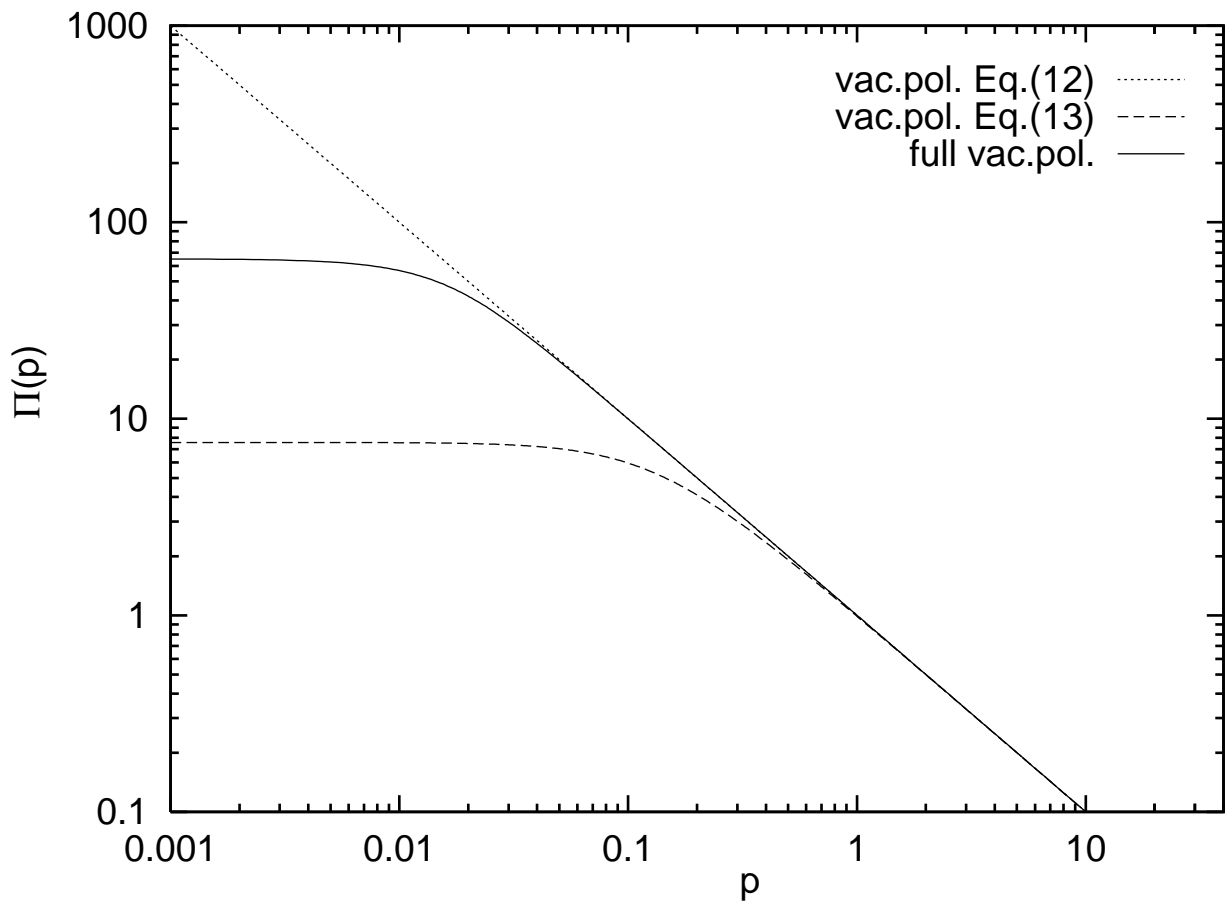
\end{document}